\documentclass[letterpaper, 10 pt, conference]{ieeeconf}  

\IEEEoverridecommandlockouts                              
\usepackage[english]{babel}
\usepackage[utf8]{inputenc}
\usepackage[T1]{fontenc}

\usepackage{amsmath} 

\usepackage{graphicx}
\graphicspath{ {./graphics/} }

\title{\LARGE \bf
Prediction of Delirium Risk in Mild Cognitive Impairment Using Time-Series data, Machine Learning and Comorbidity Patterns - A Retrospective Study
}

\author{Santhakumar Ramamoorthy$^{1}$,
        Priya Rani$^{2}$,
        James Mahon$^{3}$,
        Glenn Mathews$^{4}$, 
        Shaun Cloherty$^{5}$,
        Mahdi Babaei$^{6}$
}

\begin{document}

\maketitle
\thispagestyle{empty}
\pagestyle{empty}

\begin{abstract}

Delirium represents a significant clinical concern characterized by high morbidity and mortality rates, particularly in patients with mild cognitive impairment (MCI). This study investigates the associated risk factors for delirium by analyzing the comorbidity patterns relevant to MCI and developing a longitudinal predictive model leveraging machine learning methodologies. A retrospective analysis utilizing the MIMIC-IV v2.2 database was performed to evaluate comorbid conditions, survival probabilities, and predictive modeling outcomes. The examination of comorbidity patterns identified distinct risk profiles for the MCI population. Kaplan-Meier survival analysis demonstrated that individuals with MCI exhibit markedly reduced survival probabilities when developing delirium compared to their non-MCI counterparts, underscoring the heightened vulnerability within this cohort. For predictive modeling, a Long Short-Term Memory (LSTM) ML network was implemented utilizing time-series data, demographic variables, Charlson Comorbidity Index (CCI) scores, and an array of comorbid conditions. The model demonstrated robust predictive capabilities with an AUROC of 0.93 and an AUPRC of 0.92. This study underscores the critical role of comorbidities in evaluating delirium risk and highlights the efficacy of time-series predictive modeling in pinpointing patients at elevated risk for delirium development.

\end{abstract}

\section{INTRODUCTION}

Delirium is a clinical condition characterized by acutely impaired attention and consciousness, commonly occurring in older patients in healthcare settings. Older age, comorbid medical conditions, higher burden of acute illness, and the use of sedative and analgesic medications are significant factors associated with an increased risk for delirium [1, 2].  Delerium is commonly observed in intensive care units (ICUs) and is associated with heightened mortality rates, prolonged hospitalizations, and long-term cognitive deficits [3, 4]. Among ICU patients worldwide, delirium is estimated to affect up to 64\% during hospitalization [5, 6]. The frequency and occurrence of delirium differ among various populations and age brackets, with data suggesting a prevalence of 0.4\% in the overall population, rising to 1\% in those aged 55 years and older [7]. The occurrence of postoperative delirium (POD) is a frequent complication after surgical procedures, with documented occurrence rates varying from 13.2\% to 41.7\% [8].  Alongside cognitive manifestations, delirium is linked to disturbances in sleep-wake cycles [9]. Individuals experiencing delirium suffer from exacerbated morbidity and mortality rates, as well as heightened distress. The underlying causes of delirium are complex and poorly understood, which makes it both difficult to detect and prevent.  

Mild cognitive impairment (MCI) is a transitional stage between normal age-related cognitive decline and more severe cognitive impairment associated with conditions such as Alzheimer's disease (AD). It is characterized by cognitive deficits that are noticeable but not severe enough to interfere significantly with daily activities [10]. The presence of MCI has been associated with an increased vulnerability to delirium, especially during periods of acute stress, such as hospitalization or surgery [11]. Studies indicate that individuals with MCI may exhibit alterations in inflammatory processes that predispose them to delirium [12]. Furthermore, the cognitive deficits associated with MCI can lead to a higher likelihood of developing delirium when faced with relatively mild medical stressors [13]. Given the multifactorial nature of delirium, individualized, comprehensive strategies are essential to effectively prevent, identify, and manage this complex condition [14]. MCI is a critical risk factor for delirium. The interplay between cognitive impairment and delirium underscores the need for vigilant screening and management strategies in clinical settings to mitigate the risks associated with delirium and its potential long-term consequences on cognitive health.

Despite growing interest in understanding the relationship between MCI  and delirium, existing studies have largely focused on incidence rates or progression to dementia, without leveraging longitudinal, time-series data to assess survival outcomes or predict delirium risk over time. To address this critical gap, the present study proposes a comprehensive approach with three key objectives: (1) to identify key risk factors associated with the onset of delirium in individuals with MCI by analyzing comorbidity profiles; (2) to conduct a survival analysis comparing the incidence of delirium between MCI and non-MCI patient groups; and (3) to develop a time-series–based longitudinal predictive model capable of predicting delirium risk over time. To the best of our knowledge, no prior work has systematically applied time-series modeling to comorbidity trajectories for delirium prediction in this population, making this study both novel and timely.

\begin{figure*}[thpb]
  \centering
  \framebox{\parbox{7in}{
    \includegraphics[width=\textwidth]{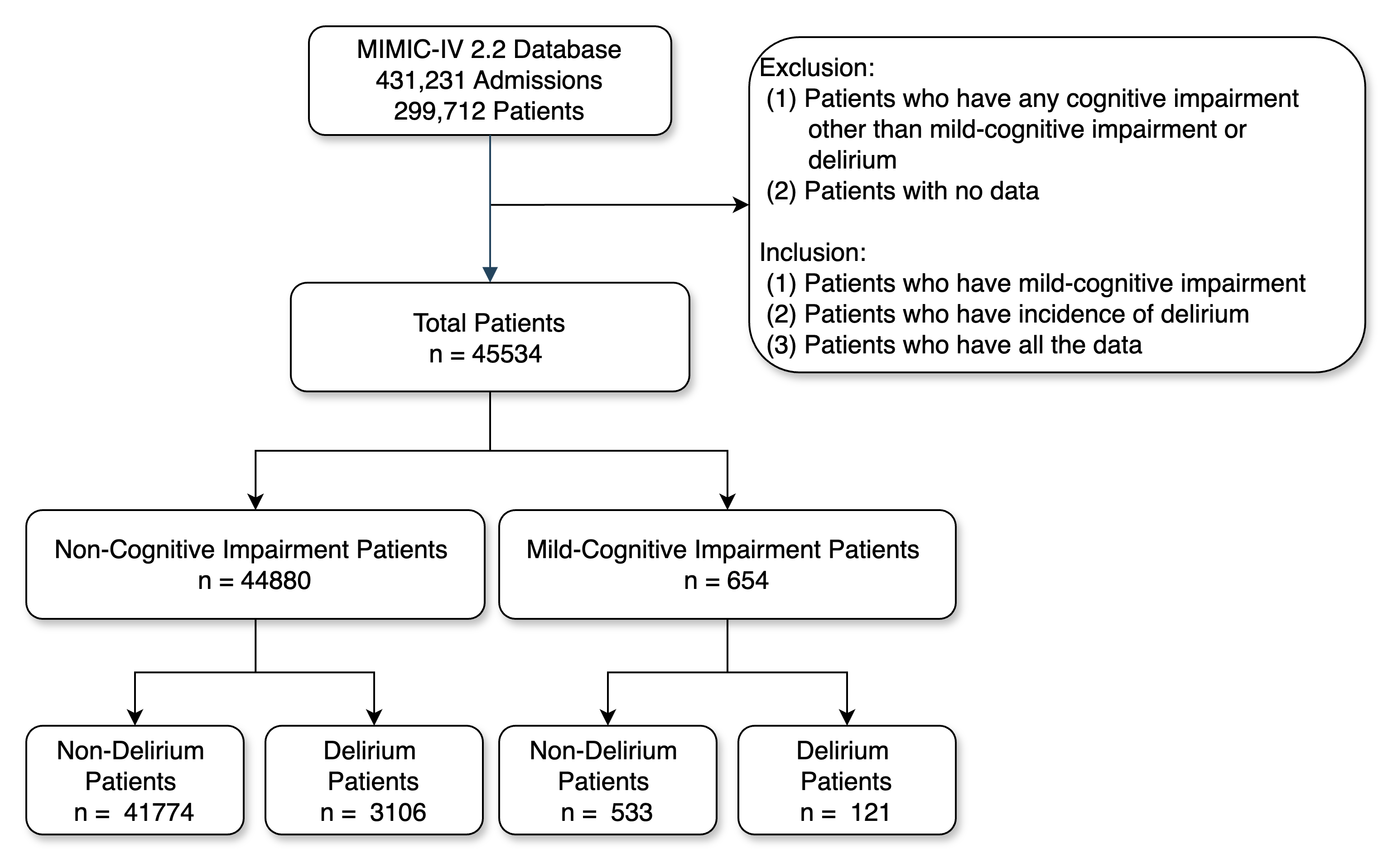}
  }}
 
  \caption{Study cohort selection flowchart from the MIMIC-IV v2.2 database. The flowchart outlines the inclusion and exclusion criteria used to identify the study population, the population was stratified into MCI and non-MCI cohorts, and each group was further divided based on the presence or absence of delirium}
  \label{fig:Flow Chart}
\end{figure*}

\section{Methods}

\begin{table*}[thpb]
    \centering
    \caption{Summary of Variables included in the Delirium prediction models}
    \label{table:Variable Data}
    \renewcommand{\arraystretch}{1.5} 
    \begin{tabular}{p{0.95\textwidth}} 
        \hline
        \textbf{Demographic data} \\
        \hspace{1em} Age, gender \\
        \textbf{Diseases diagnosis data} \\
        \hspace{1em} Total International Classification of Diseases (ICD) - coded diagnoses per admission* \\
        \textbf{Comorbidity Condition data} \\
        \hspace{1em} 
        Myocardial Infarction, Congestive Heart Failure,
        Peripheral Vascular Disease, Cerebrovascular Disease, 
        Chronic Pulmonary Disease,
        \\
        \hspace{1em}
         Rheumatic Disease, Peptic Ulcer Disease, Mild Liver Disease, 
         Diabetes Without Comorbidities, Diabetes With Comorbidities,
        \\
        \hspace{1em}
        Paraplegia, Renal Disease, Malignant Cancer, 
        Severe Liver Disease, Metastatic Solid Tumor,
        Charlson Comorbidity Index.
   
        \hspace{1em}
        \\
        \hline
        \\
    \end{tabular}
    
    \vspace{0.5em} 
    \begin{minipage}{0.95\textwidth}
    \small
    \textit{Note.} * The variable total International Classification of Diseases (ICD) - coded diagnoses per admission refers to the cumulative count of all unique ICD-coded diagnosed diseases per admission which varies for each patients
    \end{minipage}
\end{table*}

\begin{table*}[thpb]
    \centering
    \caption[]{Characteristics of MCI and Non-MCI study population.}
    \label{table: Characteristics of MCI study population table}
    \renewcommand{\arraystretch}{1.5} 
    \begin{tabular}{p{3.5cm} p{3.5cm} p{3.5cm} p{3.5cm} p{.5cm}}
    \hline
    \ & \textbf{Total} &\textbf{No MCI} &  \textbf{MCI} & \\
    \cline{2-5} 
    & \textbf{\ 45534(100\%)} & \textbf{44880(98.56\%)} & \textbf{654(1.44\%)}   \\
    \hline
     Age (IQR*), years & 79 (73-84) & 79 (73-84) & 80 (74-85) \\
     Male & 21226 (46.62\%) & 20933 (46.64\%) & 293 (44.80\%) \\
     Female & 24308 (53.38\%) & 23947 (53.36\%)& 361 (55.20\%)  \\
    \hline
    \\
    \end{tabular}
    \vspace{0.5em} 
    \begin{minipage}{0.95\textwidth}
    \small
    \textit{Note.} *The Age values are reported as medians with interquartile ranges (IQR).
    \end{minipage}
\end{table*}



\subsection{Ethical review}

The data collected by the Medical Information Mart for Intensive Care IV (MIMIC-IV) version 2.2 database [15, 16] follows recognized ethical guidelines and meets both institutional and federal compliance requirements. Due to its retrospective design, lack of direct patient involvement, and the use of a thorough de-identification process that aligns with Safe Harbor criteria outlined by the Health Insurance Portability and Accountability Act (HIPAA), the dataset does not require approval from an institutional review board (IRB). MIMIC-IV provides a modern electronic health record dataset that covers ten years of hospital admissions from 2008 to 2019, thus offering extensive and de-identified patient information for research purposes. The authors have completed the necessary coursework and personal training evaluations required by the Massachusetts Institute of Technology Affiliates (Record ID: 62569697), ensuring compliance with ethical standards for research involving de-identified health information.

\subsection{Study Population}

The study population was derived from the Medical Information Mart for Intensive Care IV (MIMIC-IV) version 2.2 database. After applying the selection criteria, a total of 45,534 patients were included in the analysis. This group excluded individuals diagnosed with any of the following cognitive impairment conditions based on International Classification of Diseases, 9th Revision (ICD-9) and 10th Revision (ICD-10) codes: 290.0, 290.1, 290.2, 290.3, 290.4, 290.8, 290.9, 291.1, 291.2, 292.82, 294.0, 294.10, 294.11, 294.20, 331.0, 331.82, 331.11, 331.19, A81.00, E710, E75.2, E75.23, E75.29, F01, F02, F03, F04, F10.26, F10.27, F10.96, F10.97, F13.26, F13.27, F13.96, F13.97, F18.27, F18.97, G10, G20, G30, G23.1, G31, G31.85, R41. The data was divided into two groups, Non-MCI Group: A total of 44880 patients were classified as having no mild cognitive impairment (MCI), within the group, 3106 patients were identified with delirium using ICD codes F05, 293.0 and 292.81. MCI Group: The mild cognitive impairment (MCI) group consisted of 654 patients identified based on the ICD-9 code 331.83 and the ICD-10 code G31.84. Among them, 121 patients were further classified into the delirium group, based on the same ICD codes F05, 293.0, and 292.81. (Fig.~\ref{fig:Flow Chart}) presents a comprehensive overview of the patient selection process. The complete list of ICD-9 and ICD-10 codes used to define inclusion and exclusion criteria is provided in Appendix section Table~\ref{table:appendix_icd_combined}.

\subsection{Identification of Key Risk Factors through Comorbidity Profiling }
To identify relevant comorbidity-related risk factors for delirium in patients with Mild Cognitive Impairment (MCI), the patients were divided into two groups: those with a diagnosis of Mild Cognitive Impairment (MCI; 654 patients) and those without MCI (44880 patients). Within both the MCI and non-MCI groups, individuals were further categorized into delirium and non-delirium subgroups using corresponding diagnostic codes<add appendix reference>. Specifically, there were 121 patients with delirium in the MCI group and 3106 patients with delirium in the non-MCI group. Comorbidity profiles were established for all groups. The specific comorbidities identified in the study included myocardial infarction, congestive heart failure, peripheral vascular disease, cerebrovascular disease, chronic pulmonary disease, rheumatic disease, peptic ulcer disease, mild liver disease, diabetes (with and without comorbidity), paraplegia, renal disease, malignant neoplasms, severe liver disease, and metastatic solid tumors. Descriptive statistics were used to calculate the proportion of patients with each comorbid condition within each group. The association between comorbidities and group membership ( MCI vs. non-MCI and delirium vs. non-delirium) was evaluated using Pearson’s Chi-square test of independence. The test determined whether observed differences in proportions were statistically significant, with a threshold of p < 0.05 considered significant. Confidence intervals (95\%) for proportions were calculated using the Wald method to quantify uncertainty. This analysis aimed to highlight which comorbidities were more prevalent in patients with MCI who developed delirium, providing insights into potential risk factors for targeted prediction.

\subsection{Kaplan-Meier Survival Analysis for Delirium Onset}
To identify relevant comorbidity-related risk factors for delirium in patients with Mild Cognitive Impairment (MCI), a Kaplan-Meier survival analysis was conducted to compare the incidence and timing of delirium onset between MCI and non-MCI cohorts. The MCI cohort included 121 patients with delirium, while the non-MCI cohort included 3,106 patients with delirium. The survival time was defined as the duration (in months) from hospital admission to the first recorded diagnosis of delirium. Patients without a delirium diagnosis were treated as right-censored at the time of discharge or at the end of the study period. Both the MCI and non-MCI groups were analyzed independently, and survival curves were generated to visualize the cumulative probability of remaining delirium-free over time. Survival probabilities were estimated at defined monthly intervals, and the 95\% confidence intervals (CIs) for each time point were calculated using Greenwood’s formula. To statistically compare the survival distributions between the two groups, the log-rank test was applied. A p-value threshold of p < 0.05 was used to determine statistical significance in survival differences. The analysis aimed to evaluate whether individuals with MCI were at higher risk of developing delirium over time compared to those without MCI.

\subsection{Development of a Time-Series Predictive Model for Delirium Onset}
\begin{figure}[thpb]
  \centering
  \framebox{\parbox{3.2in}{
    \includegraphics[width=8.2cm, height=2cm]{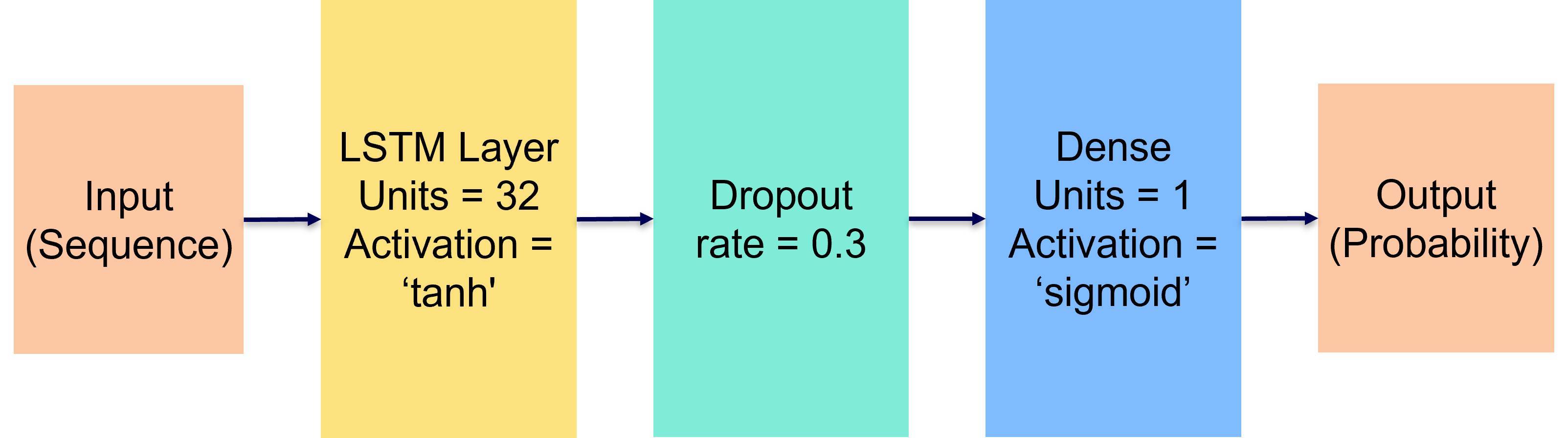}
  }}

  \caption{A Lightweight LSTM-Based Architecture for Predicting Delirium Onset Using Sequential Data}
  \label{fig:LSTM-Based Architecture}
\end{figure}

To dynamically predict the onset of delirium over time, a longitudinal time-series model was developed using a Long Short-Term Memory (LSTM) neural network architecture [17]. The model was designed to capture sequential patterns in patient data across hospital admissions and to estimate individualized delirium risk for patients diagnosed with Mild Cognitive Impairment (MCI). The proposed architecture consist of a single LSTM layer to capture temporal dependencies from sequential clinical data, followed by a dropout layer to prevent overfitting. A dense layer with sigmoid activation is used for predicting delirium onset (Fig.~\ref{fig:LSTM-Based Architecture}). This lightweight design ensures computational efficiency while maintaining predective performance suitable for real-time clinical application. The input dataset was constructed from structured electronic health records, including demographic variables (age, gender), total ICD-coded diagnoses per admission, comorbidity indicators (based on the Charlson Comorbidity Index) [18], and Charlson Comorbidity Index (CCI) scores (Table. I). For each patient, temporal sequences were generated to reflect their clinical trajectory over successive admissions. To address the class imbalance caused by the relatively smaller number of delirium-positive cases, SMOTENC (Synthetic Minority Over-sampling Technique for Nominal and Continuous features) [19] was applied to the flattened time-series data to upsample the minority class. Additionally, the majority (non-delirium) class was downsampled using random sampling to reduce potential model bias. The model was trained and validated using a 10-fold cross-validation approach to ensure robustness. Performance metrics included Area Under the Receiver Operating Characteristic Curve (AUROC) and Area Under the Precision-Recall Curve (AUPRC) to assess classification performance. Model calibration was evaluated using the Brier score, indicating the accuracy of probabilistic predictions.

\section{Result}

\begin{table*}[thpb]
    \centering
    \caption{Comorbidity characteristics of the study population}
    \label{table: Characteristics of the study population table}
    \renewcommand{\arraystretch}{1.5} 
    \begin{tabular}{l l l l l l l l} 
    \hline
    \textbf{Parameters}   & \multicolumn{3}{c}{\textbf{MCI Cohorts}} &  & \multicolumn{3}{c}{\textbf{Delirium Cohorts}}   \\
    \cline{2-4} 
    \cline{6-8} 
    \ & \textbf{with MCI} &  \textbf{without MCI} &    &   & \textbf{with MCI} & \textbf{without MCI}  & \\
    \cline{2-4} 
    \cline{6-8} 
    & \textbf{\( \hat{p} \) (95\%CI)} & \textbf{\( \hat{p} \) (95\%CI)} & \textbf{\textit{p}-value} &   & \textbf{\( \hat{p} \) (95\%CI)} & \textbf{\( \hat{p} \) (95\%CI)} &\textbf{\textit{p}-value} \\
    \hline
     \textbf{Peripheral Vascular Disease} & \textbf{0.086 (0.064-0.107)} & \textbf{0.114 (0.111-0.117)} & \textbf{<0.022} & & 0.132 (0.072-0.193)& 0.154 (0.141-0.167) &  0.516\\
     \textbf{Renal Disease} & \textbf{0.292 (0.257-0.327)} & \textbf{0.233 (0.229-0.237)} &   \textbf{<0.000} & & 0.405 (0.317-0.492) & 0.332 (0.315-0.348)&  0.095\\
     \textbf{Diabetes With Comorbidities} & \textbf{0.136 (0.110-0.162)} & \textbf{0.090 (0.087-0.092)} &   \textbf{<0.000} & & \textbf{0.198 (0.127-0.269)}& \textbf{0.120 (0.108-0.131)}&  \textbf{<0.010}\\
     \textbf{Metastatic Solid Tumor} & \textbf{0.035 (0.021-0.049)} & \textbf{0.069 (0.066-0.071)}&  \textbf{<0.001} & & \textbf{0.025 (0.000-0.052)}& \textbf{0.081 (0.072-0.091)}& \textbf{<0.024}\\
     Myocardial Infarction & 0.133 (0.107-0.159) & 0.147 (0.143-0.150) & 0.325 & & 0.198 (0.127-0.269) & 0.175 (0.162-0.189) & 0.517\\
     Congestive Heart Failure & 0.271 (0.237-0.305) & 0.272 (0.268-0.276) & 0.946 & & 0.405 (0.317-0.492)& 0.397 (0.380-0.415) &  0.866\\
     Cerebrovascular Disease & 0.139 (0.113-0.166) & 0.130 (0.127-0.133)&   0.479 & & 0.149 (0.085-0.212)& 0.146 (0.134-0.159) &  0.945\\
     Chronic Pulmonary Disease & 0.206 (0.175-0.237) & 0.227 (0.223-0.231)&   0.061 & & 0.256 (0.178-0.334)& 0.280 (0.265-0.296) &  0.560\\
     Rheumatic Disease & 0.057 (0.039-0.074) & 0.042 (0.040-0.044)&   0.460 & & 0.066 (0.022-0.110)& 0.048 (0.040-0.055) &  0.353\\
     Peptic Ulcer Disease & 0.009 (0.002-0.016) & 0.018 (0.017-0.019) &   0.085 & & 0.017 (0.000-0.039)& 0.024 (0.018-0.029) &  0.617\\
     Mild Liver Disease & 0.032 (0.019-0.046) & 0.040 (0.038-0.042) &   0.311 & & 0.017 (0.000-0.039) & 0.044 (0.037-0.051) &  0.143\\
     Diabetes Without Comorbidities & 0.197 (0.167-0.228) & 0.215 (0.211-0.219) &   0.274 & & 0.248 (0.171-0.325)& 0.242 (0.227-0.258)&  0.890\\
     Paraplegia & 0.034 (0.020-0.047) & 0.029 (0.027-0.030) &   0.464 & & 0.017 (0.000-0.039) & 0.027 (0.021-0.032)&  0.492\\
     Malignant Cancer &  0.121 (0.096-0.146) & 0.145 (0.142-0.148) &   0.080 & & 0.107 (0.052-0.163) & 0.169 (0.156-0.182)&  0.076\\
     Severe Liver Disease &  0.006 (0.000-0.012) & 0.012 (0.011-0.013) &   0.190 & & 0.000 (0.000-0.000) & 0.012 (0.008-0.016)&  0.221\\
    
    \hline
    \\
    \end{tabular}
    \vspace{0.5em} 
    \begin{minipage}{0.95\textwidth}
    \small
    \textit{Note.} The 95\% confidence intervals (CIs) for proportions were calculated using the Wald method, and the reported p-values were obtained using the Pearson’s Chi-square test for independence.
    \end{minipage}
\end{table*}

    

    
   
    

\subsection{Patient characteristics} 

The study population, derived from the Medical Information Mart for Intensive Care IV (MIMIC-IV) version 2.2 database, consisted of three groups: patients without mild cognitive impairment (non-MCI), patients with mild cognitive impairment (MCI), and a subset of MCI and non-MCI patients diagnosed with delirium. For the non-MCI cohort, which included a total of 44,880 patients, there was a slightly higher proportion of females compared to males. Specifically, 20,933 patients (46.64\%) were male, while 23,947 patients (53.36\%) were female. The median age in this cohort was 79, with an interquartile range (IQR) of (73 – 84) years, representing an older population overall. Table.~\ref{table: Characteristics of MCI study population table} summarizes the demographic characteristics of the study population, comparing individuals with and without Mild Cognitive Impairment (MCI). The Mild Cognitive Impairment (MCI) cohort consisted of 654 patients, with a slightly greater female representation. Among this group, 293 patients (44.80\%) were male, and 361 patients (55.20\%) were female. The median age of the MCI cohort was slightly higher at 80, with an interquartile range (IQR) of (74 – 85)  years.  Furthermore, a large percentage of patients experienced multiple hospital stays throughout the study period, highlighting the ongoing healthcare demands of this group. These frequent hospitalizations reflect the complex and chronic medical conditions common among the cohort, which require persistent medical care and coordinated support.


\subsection{Comorbidity Patterns in MCI and Non-MCI Cohorts}
The analysis identified distinct comorbidity patterns in the MCI cohort compared to the non-MCI cohort, despite the smaller population size of the MCI group. In the non-MCI cohort, metastatic solid tumors were more commonly observed compared to the MCI cohort (p < 0.001). Conversely, peripheral vascular disease (p < 0.022), diabetes with comorbidities (p < 0.000), and renal disease (p < 0.000) were observed at higher proportions in the MCI cohort relative to the non-MCI group. Table.~\ref{table: Characteristics of the study population table} shows the comorbidity characteristics of the study population. These findings suggest differences in the clinical characteristics of patients with and without mild cognitive impairment.

\subsection{Comorbidity Patterns in Delirium and Non-Delirium Groups}
Within the MCI and non-MCI cohorts, additional differences were observed between patients with and without delirium. Diabetes with comorbidities was more frequent in the delirium group compared to the non-delirium group (\textit{p} < 0.010). In contrast, metastatic solid tumors were less prevalent in the delirium group than in the non-delirium group ( \textit{p} < 0.024) as shown in Table.~\ref{table: Characteristics of the study population table}. These findings further emphasize the heterogeneity within the MCI population, with specific comorbidities, such as diabetes with comorbidities, being more strongly associated with delirium, while others, like metastatic solid tumors, being less common in this subgroup.

\subsection{Kaplan-Meier Survival Analysis for Developing Delirium} 
\begin{figure}[thpb]
      \centering
      \framebox{\parbox{3.2in}{
        \includegraphics[width=8.2cm, height=6cm]{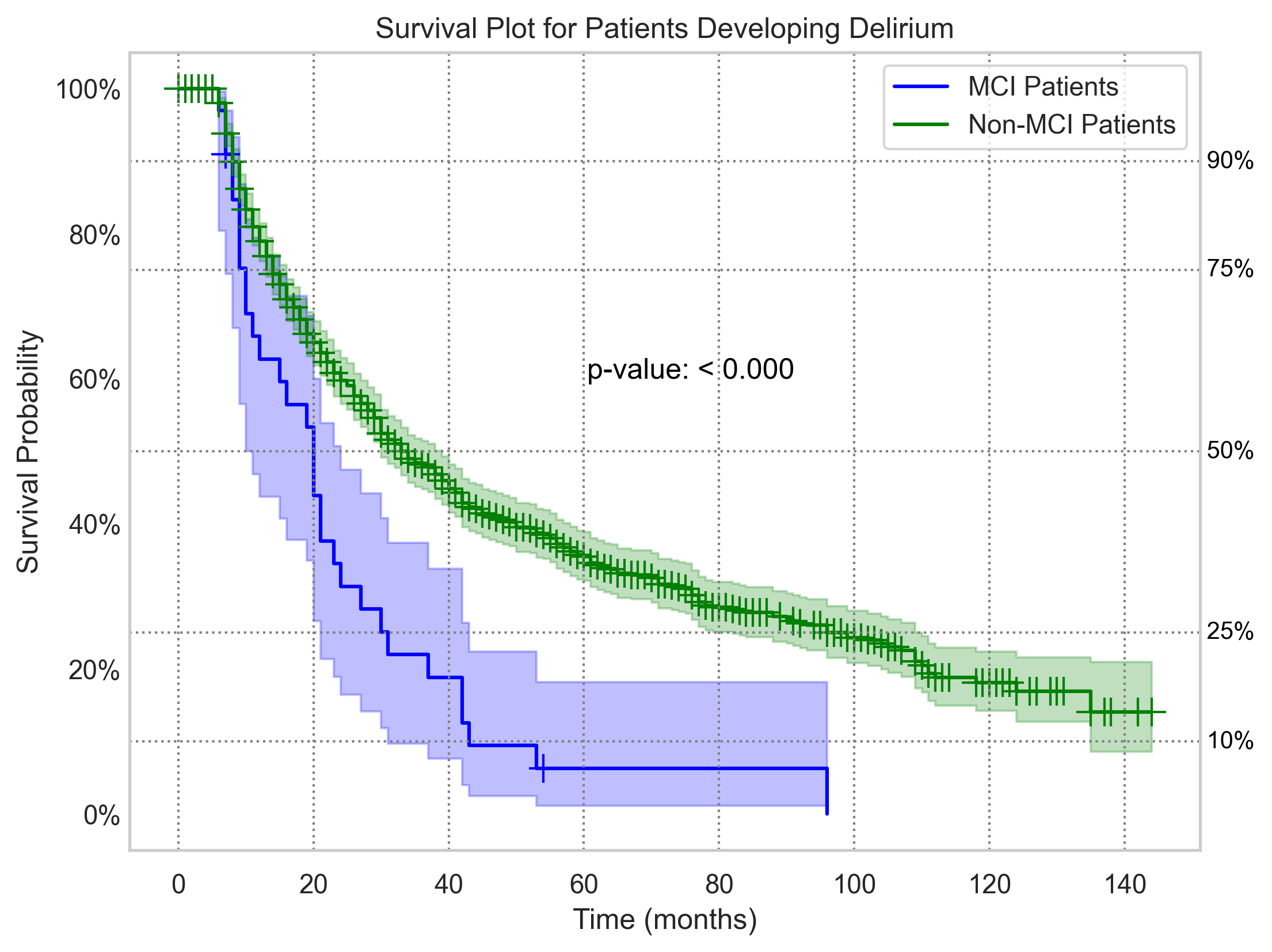}
      }}
    
      \caption{The Kaplan-Meier survival curves comparing the time to delirium onset between the Non-MCI and MCI cohorts. The plot shows a significantly lower survival probability (delirium-free duration) over time in the MCI cohort, indicating a higher risk of developing delirium compared to the Non-MCI group.}
      \label{fig:Kaplan-Meier curves}
   \end{figure}

The Kaplan-Meier survival analysis demonstrated notable differences in survival probabilities for developing delirium between the MCI and non-MCI cohorts (log-rank test, \textit{p} < 0.000). In the MCI cohort, survival probabilities declined rapidly over time, starting at 96.97\% (95\% CI: 80.37–99.57) at 6 months. By 12 months, the survival probability had dropped to 62.70\% (95\% CI: 43.74–76.83), and at 24 months, it further declined to 31.35\% (95\% CI: 16.44–47.46). In contrast, the non-MCI cohort exhibited higher survival probabilities throughout the same period. At 6 months, the survival probability was 97.96\% (95\% CI: 96.86–98.68). By 12 months, the survival probability was 78.92\% (95\% CI: 76.19–81.37), and by 24 months, it decreased to 59.72\% (95\% CI: 56.46–62.82) as shown in (Fig.~\ref{fig:Kaplan-Meier curves}). These results highlight a steeper decline in survival probabilities within the MCI cohort, indicating a higher susceptibility to developing delirium over time compared to the non-MCI cohort.

\subsection{Performance of machine learning model}
\begin{figure}[thpb]
      \centering
      \framebox{\parbox{3.2in}{
        \includegraphics[width=8.2cm, height=6cm]
        {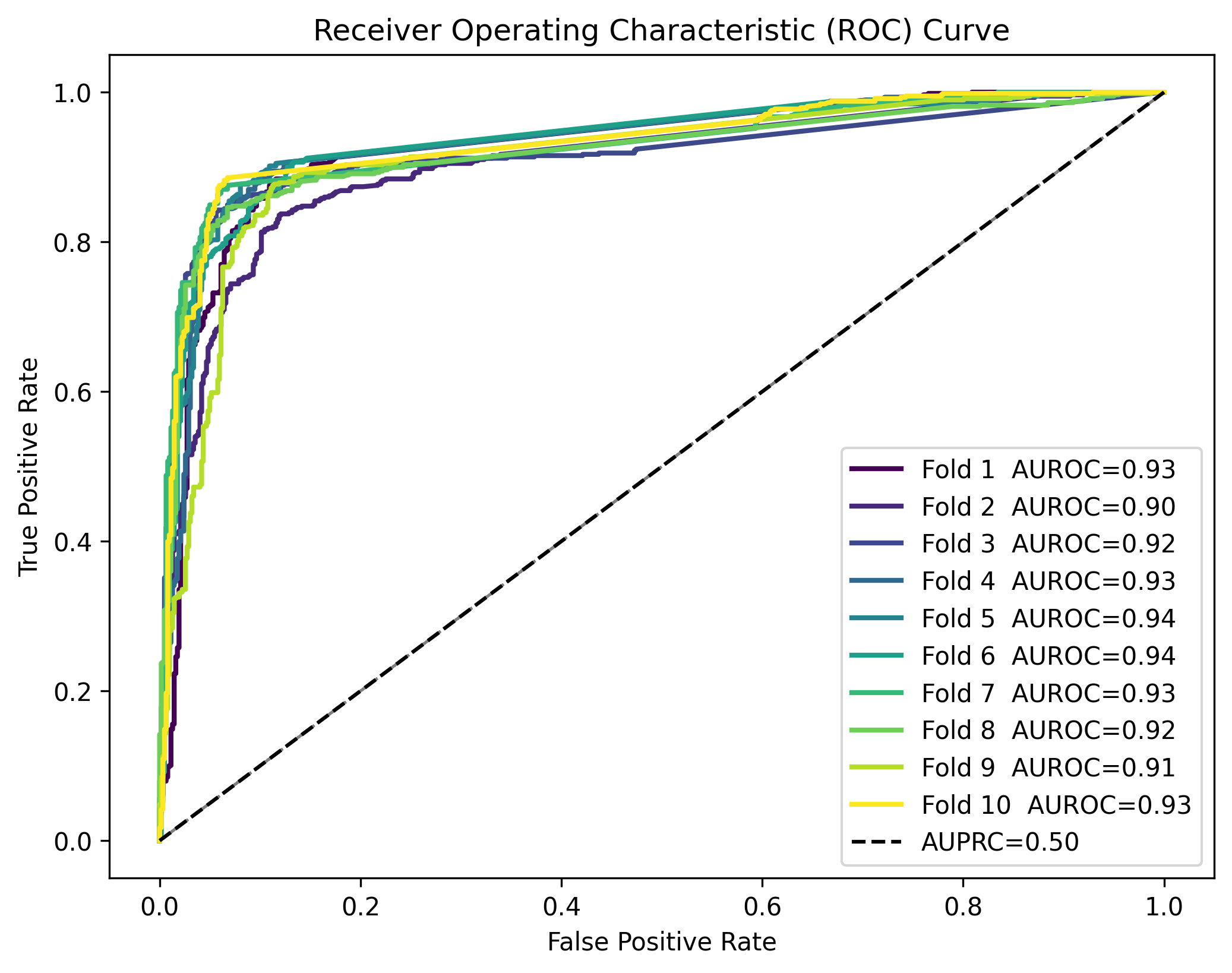}
      }}
    
      \caption{The Receiver Operating Characteristic (ROC) curve for the validation cohort, illustrating the model’s ability to discriminate between patients who developed delirium and those who did not. A high area under the curve (AUC) indicates strong predictive performance..}
      \label{fig:ROC curve}
   \end{figure}

A Long Short-Term Memory (LSTM)machine learning model was developed to predict the occurrence of delirium using time-series data from patient admissions. The model incorporated both numerical and categorical features to capture temporal and clinical patterns associated with delirium risk. The LSTM model achieved strong performance, with an area under the receiver operating characteristic curve (AUROC) of 0.93 (95\% CI: 0.91–0.94) and an area under the precision-recall curve (AUPRC) of 0.92 (95\% CI: 0.92–0.95) (Fig.~\ref{fig:ROC curve}). These metrics indicate a high level of accuracy and reliability in predicting delirium risk based on the selected features and modeling approach.

\section{DISCUSSION}

This study elucidates significant differences in comorbidity profiles, survival probabilities, and predictive modeling outcomes for delirium in patients with mild cognitive impairment (MCI) and highlights the heterogeneity of this population compared to non-MCI counterparts. The findings underscore the interplay between cognitive impairment, comorbidities, and delirium risk, offering critical insights into clinical decision-making and risk stratification. A recent study indicates that adults who are experiencing MCI are at an increased risk of developing delirium [20, 21]. Also, patients with substantial comorbidities may have an elevated risk of developing delirium [22, 23, 24]. 

This study identified several statistically significant differences in comorbidity patterns between patients with Mild Cognitive Impairment (MCI) and those without, as well as between patients with and without delirium across both MCI and non-MCI cohorts. In comparison to non-MCI patients, those with MCI had a significantly lower prevalence of peripheral vascular disease (8.6\%, 95\% CI: 6.4--10.7 vs. 11.4\%, 95\% CI: 11.1--11.7; \textit{p} < 0.022) and metastatic solid tumors (3.5\%, 95\% CI: 2.1--4.9 vs. 6.9\%, 95\% CI: 6.6--7.1; \textit{p} < 0.001). Conversely, the MCI cohort showed a significantly higher prevalence of renal disease (29.2\%, 95\% CI: 25.7--32.7 vs. 23.3\%, 95\% CI: 22.9--23.7; \textit{p} < 0.000) and diabetes with comorbidities (13.6\%, 95\% CI: 11.0--16.2 vs. 9.0\%, 95\% CI: 8.7--9.2; \textit{p} < 0.000).  Across both MCI and non-MCI cohorts, patients who developed delirium had a higher prevalence of diabetes with comorbidities (19.8\%, 95\% CI: 12.7--26.9 vs. 12.0\%, 95\% CI: 10.8--13.1; \textit{p} < 0.010). These findings indicate that specific comorbidities, such as diabetes with comorbidities, may serve as potent risk factors for delirium within the MCI cohort [25].


Kaplan-Meier survival analysis revealed a significantly steeper decline in survival probabilities for developing delirium among MCI patients compared to non-MCI patients. At 24 months, the survival probability for the MCI cohort dropped to 31.35\%, compared to 59.72\% in the non-MCI cohort (log-rank test, \textit{p} < 0.000). This underscores the heightened vulnerability of cognitively impaired patients to delirium, which may be driven by their distinct comorbid profiles and underlying pathophysiological processes. The delirium group within the MCI cohort also exhibited poorer survival probabilities compared to their non-MCI counterparts, further emphasizing the clinical burden of delirium in this population. 

The robust performance of the LSTM-based prediction model (AUROC: 0.93; AUPRC: 0.92) validates the integration of time-series health records, including demographic data, Charlson Comorbidity Index (CCI) scores, and detailed comorbidity profiles, as critical components for accurate delirium risk stratification. The model's ability to effectively capture temporal and clinical patterns in high-risk populations represents a significant advancement in predictive analytics for complex patient cohorts.

\section{LIMITATIONS}

Despite its strengths, this study has several limitations. The retrospective design and exclusive use of structured electronic health record (EHR) data may limit the generalizability of the findings. The predictive model was developed and validated using the MIMIC-IV database, which contains data from a single academic medical center in the United States. This may restrict the applicability of the model across different geographic regions and patient populations with varying demographic and ethnic characteristics. Furthermore, the relatively small sample size of the MCI cohort may have reduced the statistical power to detect clinically meaningful differences. Future research should explore the integration of diverse data.

\section{CONCLUSIONS}

In this study, the interplay between mild cognitive impairment (MCI), comorbidities, and the risk of delirium was systematically evaluated, along with the development of a time-series predictive model. Analysis in this study revealed that distinct comorbidity profiles significantly influenced delirium risk in MCI patients. Furthermore, a Long Short-Term Memory (LSTM) model was successfully implemented, confirming the utility of time-series data and machine learning in predicting delirium risk. The model demonstrated robust performance in identifying high-risk patients, thereby supporting the integration of advanced analytics into clinical workflows.

\addtolength{\textheight}{-12cm}   



\section*{APPENDIX}

\subsection{Inclusion and Exclusion Criteria ICD Codes}
This appendix provides a summary of the International Classification of Diseases (ICD-9 and ICD-10) codes used to define the inclusion and exclusion criteria for study participants. These codes, as detailed in Table~\ref{table:appendix_icd_combined}, were selected to identify cases of mild cognitive impairment (MCI) and delirium while excluding other cognitive and neurodegenerative conditions that may confound the analysis.

\renewcommand{\thetable}{A.\arabic{table}}
\setcounter{table}{0}
\begin{table*}[t]
\centering
\caption{ICD Codes and Conditions Used for Study Inclusion and Exclusion (ICD-9 and ICD-10)}
\label{table:appendix_icd_combined}

\renewcommand{\arraystretch}{1.3}
\small
\begin{tabular}{p{7cm} p{9.5cm}} 
\hline
\textbf{ICD Code (Version)} & \multicolumn{1}{c}{\textbf{Condition Name}} \\
\hline
290.0–290.9 (ICD-9) & Senile and presenile dementias \\
291.1, 291.2 (ICD-9) & Alcohol-induced persisting amnestic disorder / dementia \\
292.81 (ICD-9) & Drug-induced delirium \\
292.82 (ICD-9) & Drug-induced persisting dementia \\
293.0 (ICD-9) & Delirium, not otherwise specified (NOS) \\
294.0–294.20 (ICD-9) & Persistent mental disorders due to other conditions \\
331.0 (ICD-9) & Alzheimer’s disease \\
331.11, 331.19 (ICD-9) & Pick’s disease / Frontotemporal dementia \\
331.82 (ICD-9) & Dementia with Lewy bodies \\
331.83 (ICD-9) & Mild Cognitive Impairment \\
A81.00 (ICD-10) & Creutzfeldt–Jakob disease \\
E71.0, E75.2, E75.23, E75.29 (ICD-10) & Metabolic and genetic neurodegenerative disorders \\
F01–F04 (ICD-10) & Vascular and unspecified dementias \\
F10.26–F18.97 (ICD-10) & Substance-induced persisting dementia and amnestic disorders \\
G10 (ICD-10) & Huntington’s disease \\
G20 (ICD-10) & Parkinson’s disease \\
G30 (ICD-10) & Alzheimer’s disease \\
G23.1, G31, G31.85 (ICD-10) & Other degenerative diseases of the nervous system \\
G31.84 (ICD-10) & Mild Cognitive Impairment \\
R41 (ICD-10) & Symptoms involving cognitive function and awareness \\
F05 (ICD-10) & Delirium due to known physiological condition \\

\hline

\\
\end{tabular}
    \vspace{0.5em} 
    \begin{minipage}{0.95\textwidth}
    \small
    \textit{Note.} The table above outlines the ICD-9 and ICD-10 codes used to define inclusion and exclusion criteria for this study. These codes represent various cognitive impairment and neurodegenerative conditions [26].
    \end{minipage}
\end{table*}




\end{document}